\documentclass[
    ,final            
  ]
  {aipproc}

\layoutstyle{6x9}
\usepackage{graphicx}

\def    \be             {\begin{equation}}
\def    \ee             {\end{equation}}
\def    \ba             {\begin{eqnarray}}
\def    \ea             {\end{eqnarray}}

\def    \=              {\;=\;}
\def    \frac           #1#2{{#1 \over #2}}

\def    \mev            {\mbox{$\mathrm{MeV}$}}
\def    \gev            {\mbox{$\mathrm{GeV}$}}
\def\mub {\mbox{$\mu\mathrm{b}$}}

\def    \pt             {\mbox{$p_T$}}
\def    \et             {\mbox{$E_T$}}
\def    \ptb             {\mbox{$p_T^b$}}

\newcommand     \MSB            {\ifmmode {\overline{\rm MS}} \else 
                                 $\overline{\rm MS}$  \fi}

\def    \as             {\ifmmode\alpha_s\else$\alpha_s$\fi}
\def    \asmz             {\ifmmode\alpha_s(m_Z)\else$\alpha_s(m_Z)$\fi}

\def \ss{\scriptstyle}
\def \sss{\scriptscriptstyle}
\def \lsim{\mathrel{\vcenter
     {\hbox{$<$}\nointerlineskip\hbox{$\sim$}}}}

\def\ppbar{\mbox{$p \bar{p}$}}
\def\met{$\rlap{\kern.2em/}E_T$}

\begin{document}
\begin{flushright}{CERN-PH-TH/04-210}\end{flushright}
\title{The saga of bottom production in $\ppbar$
  collisions
\footnote{Presented at the 2004 Hadron Collider Physics
 Workshop, East Lansing, MI, June 2004}
}

\author{Michelangelo L. Mangano}{
  address={TH Division, CERN, Geneva, Switzerland}
}

\begin{abstract}
I review here the history of bottom quark cross section measurements
and theoretical predictions. Starting form the early days of UA1, and
going through the sequence of the large excesses reported during run~0
and~I at
the Tevatron by CDF and D0, I summarize how both
data and theory have evolved in time, thanks to improved 
experimental techniques, more data, and improved control over the main
ingredients of the theoretical calculations.
I conclude with the discussion of the preliminary data from run~II,
which appear to finally give a satisfactory picture of the data vs
theory comparison.
\end{abstract}

\maketitle

\section{Introduction}
The study of events with bottom quarks has led in the past 10 years to
some of the most important Tevatron results: the discovery and study
of the top quark, the appreciation of the colour-octet-mediated
quarkonium production mechanisms, as well as general results in
b-hadron physics (spectroscopy, lifetimes, mixing, $\sin^2 2\beta$)
These results have been obtained while both CDF and D0 were reporting
factor-of-3 discrepancies between observed and predicted $b$-hadron
cross-sections. To claim that we need to understand $b$ production in
order to make new discoveries is therefore a bit exagerated: important
discoveries should be able to stand on their feet without appealing to
the prediction of a QCD calculation.  Nevertheless, lack of confidence
in the ability to describe the properties of events containing $b$
quarks, in addition to raising doubts over the general applicability
of perturbative QCD in hadronic collisions, does limit our potential
for the observation of new dynamical regimes (e.g. small-$x$
physics~\cite{Collins:1991ty}-\cite{Jung:2001rp}) or for the discovery
of new phenomena (e.g. Supersymmetry~\cite{Berger:2000mp}).  In some
cases, the existing measurements challenge the theory in ways which go
beyond simple overall normalization issues, pointing at effects which
are apparently well beyond reasonable theoretical systematics: this is
the case of recent CDF studies, which detected anomalies in both rates
and properties of events with secondary vertices and soft
leptons~\cite{Acosta:2001ct}.  It cannot be contested, therefore, that
the study of $b$ production properties should be one of the main
priorities for Run~II at the Tevatron, with implications which could
go beyond the simple study of QCD.

Starting from the situation as it developed during the early Tevatron
runs, I will review here the progress in the theoretical predictions.
More details on the historical evolution of the cross section
measurements can be found in~\cite{mlmtalk}, as well as
in~\cite{Frixione:1994nb,Frixione:1997ma}, which also review the status of
fixed-target heavy quark studies. For a recent review including
$\gamma\gamma$ and $ep$ data as well, see~\cite{Cacciari:2004ur}.
I will then present the
implications of the preliminary results from Run II. Their
complete theoretical analysis  is contained
in~\cite{Cacciari:2003uh}.

\section{Review of run~0 and run~I results}
The prehistory of $b$ cross-section measurements in hadronic
collisions starts with UA1 at the S$\bar{p}p$S ($\sqrt{S}=630$~GeV)
collider~\cite{Albajar:1987iu}. The data were compared with
theoretical predictions~\cite{Nason:1988xz,Nason:1989zy},
showing good agreement, within the rather large ($\pm 40\%$)
theoretical uncertainty. ``Theory'', in those days, already meant a full NLO
QCD calculation~\cite{Nason:1988xz,Nason:1989zy}, 
including all mass effects, state-of-art NLO PDF
fits~\cite{Diemoz:1987xu}, and $b\to B$ non-perturbative fragmentation
functions parameterized according to~\cite{Peterson:1982ak}, with a
parameter $\epsilon=0.006$ extrapolated from fits~\cite{Chrin:1987yd} to charm
fragmentation data in $e^+e^-$, using the relation
$\epsilon_b=\epsilon_c \times (m_c/m_b)^2$.
At the beginning only predictions for total cross-sections and
inclusive \ptb\ spectra were available. Later on, more exclusive
calculations were performed, allowing for the application of general
cuts to the final states, as well as for the study of correlations
between the $b$ and 
$\bar{b}$~\cite{Mangano:1991jk}\footnote{For lack of time, I
  will however focus my attention in this presentation on inclusive \pt\
  spectra.}.  
 
After such a good start in UA1, 
the first published data from CDF~\cite{Abe:1992fc} appeared as a big
surprise. CDF collected a sample of $14\pm 4$ fully reconstructed
$B^{\pm} \to \psi K^{\pm}$ decays, leading to:
\be
\sigma(\ppbar \to bX; \; \ptb>11.5\gev,\; \vert y \vert<1)=
\begin{array}{ll}
\mathrm{CDF:} & 6.1\pm 1.9_{\ss stat} \pm 2.4_{\ss syst} \,\mu{\mathrm
  b} \\
\mathrm{theory:} & 1.1\pm 0.5 \, \mu{\mathrm
  b} 
\end{array}
\ee
In spite of the large uncertainties, which led to a mere 1.5$\sigma$
discrepancy, attention focused on the large data/theory=5.5 excess.
Theoretical work to explain the apparent contradiction between the
success of the NLO theory at 630~\gev\ and the disaster at 1.8~TeV
concentrated at the beginning on possible effects induced by the
different $x$ range probed at the two energies: PDF
uncertainties~\cite{Berger:1992je} and large small-$x$
effects~\cite{Collins:1991ty}-\cite{Levin:1991ry}, where $x\sim
m_b/\sqrt{S}$. In the first case
marginal fits to both data sets could be obtained at the cost of
strongly modifying the gluon density, in a way which however would not
survive the later accurate determinations of $g(x)$ from HERA. In the
second case, conflicting conclusions were reached: on one side the
first paper of~\cite{Levin:1991ry} obtained increases by factors of
3-5 due to small-$x$ effects; on the other, the analysis
of~\cite{Collins:1991ty} proved that the resummation of small-$x$
logarithms could only augment the total rate by 30\% (or less, in the
case of $g(x)$ more singular than $1/x$)\footnote{The option of very
  large small-$x$ effects being manifest only at 1.8TeV
  will be definitely ruled out several years
  later, when CDF measured~\cite{Acosta:2002qk} the $b$ cross section at
  $\sqrt{S}=630$GeV and showed that the scaling from 630 to 1.8TeV was
  consistent with the predictions of pure NLO QCD.}.

\begin{figure}[t]
\includegraphics[height=.3\textheight, clip]{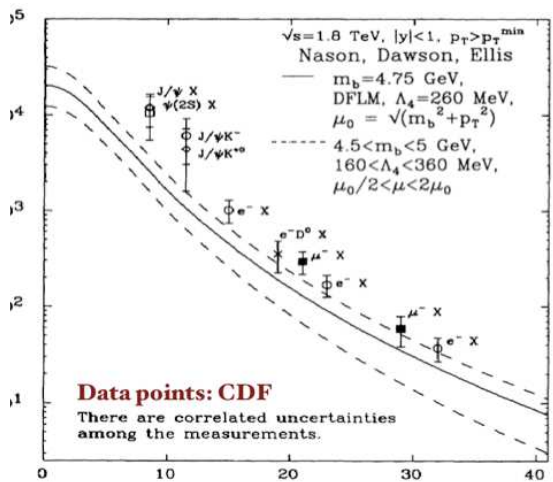}
\caption{CDF data from inclusive $\psi$, $\psi'$~\cite{Abe:1992ww}
 and lepton~\cite{Abe:1993sj} final
  states, compared to NLO QCD.}
\label{fig:cdf93}
\end{figure}

The ball was therefore back on the experimentalists' court. CDF
expanded the set of measurements, including final states with
inclusive $\psi$ and $\psi'$~\cite{Abe:1992ww} and inclusive
leptons~\cite{Abe:1993sj}, summarised in fig.~\ref{fig:cdf93}.
The measurement of the $b$ cross section from the inclusive charmonium
decays turned out later to be incorrect.
In run 0, in fact, CDF could not measure secondary vertices,
so that charmonium states from direct production and from $B$ decays
could not be separated. The extraction of a $b$ rate from these final
states was based on theoretical prejudice about the prompt
production rates, prejudice which in run~I, when the secondary
vertices started being measured by CDF, turned out to be terribly
wrong~\cite{onium}\footnote{Incidentally, this fact puts into question
  the UA1 results, which heavily relied on the $\psi$ final states and
  on explicit assumptions about the prompt charmonium rates!}. 
The data on inclusive leptons,
while high compared to the central value of the theoretical
prediction, were nevertheless 
consistent with its upper value, and in any case within
1$\sigma$. 

\begin{figure}[t]
  \includegraphics[width=.3\textheight, angle=-90]{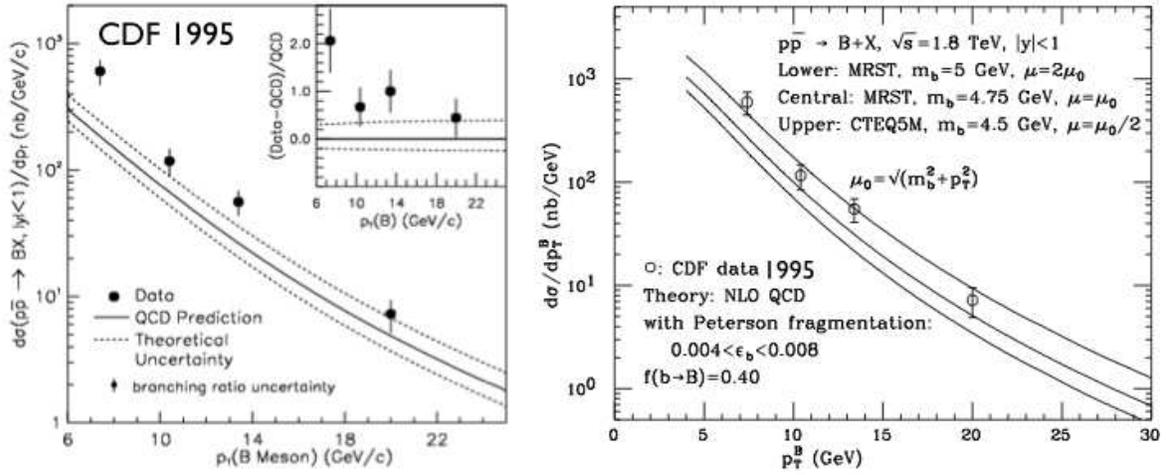}
\caption{Evolution of data/theory comparisons with improved PDF
  fits. The data on both plots are exactly the same; the theory curves
  on the left were generated with the MRSD0 set, on the right with the
  post-HERA set CTEQ5 and MRST.}
\label{fig:cdf95}
\end{figure}

Increased statistics in run~I allowed CDF to improve its measurement
of fully reconstructed exclusive decay modes, leading to the
measurements in fig.~\ref{fig:cdf95}. For this measurement CDF used
19pb$^{-1}$ of data, leading to approximately 55 $B^0\to \psi K^*$
and 125 $B^\pm \to \psi K^\pm$ decays. The cross section was still
high compared to the central value of the theoretical prediction
(data/theory=1.9$\pm$0.3), but this was already a marked improvement
over the first measurement from run~0, when this ratio was equal to 6.1! 
More explicitly, the 1995 measurement gave $\sigma(\pt(B^+)>6\gev, \vert
y\vert<1)=2.39\pm 0.54\mu$b, compared to the 1992 measurement of  $\langle\sigma(\pt(B)>9\gev, \vert
y\vert<1)\rangle=2.8\pm 1.4\mu$b (where $\langle \sigma(B) \rangle \equiv
[\sigma(B^+)+\sigma(B^0)]/2)$. Taking into account that the $b$ rate is
expected to increase by 2.7
when going from a 9~GeV to a 6~GeV threshold, the 1992 measurement
appears to be a factor of 3.2 higher than the 1995 result, consistent
with the 6.1/1.9 ratio.   
\begin{figure}[t]
  \includegraphics[width=.3\textheight, angle=-90]{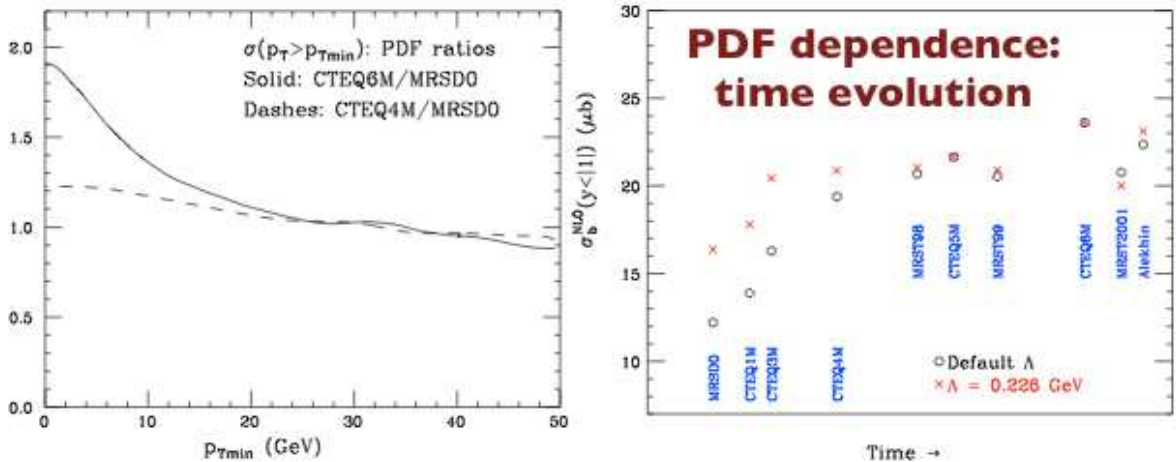}
\caption{Left: the NLO $b$-quark rate as a function of $p_{T,min}$,
  for post-HERA PDF sets CTEQ4M (\cite{Lai:1996mg}, \asmz=0.116) 
  and CTEQ6M (\cite{Pumplin:2002vw}, \asmz=0.118), normalized to the pre-HERA
  set MRSD0.  Right: total cross section for $\vert y \vert < 1$ for
  various PDF sets, distributed on the abscissa in order of increasing
  release date. The crosses correspond to the rates calculated by
  forcing $\Lambda_{\sss QCD}$ to take a value consistent with the
  LEP $\as(m_Z)$ fits ($\Lambda_{nf=5}^{2-loop}=226\mev \Rightarrow \asmz=0.118$ . }
\label{fig:pdfevol}
\end{figure}
This drop  in the
experimental cross-section was  not inconsistent with the large
statistical and systematic uncertainties of the 1992 measurement, but
somehow the common belief that theory was way off had already stuck. It
is also worth noting that the same data, when compared to theoretical
predictions obtained a couple of years later using the same QCD
calculations, but up-to-date sets of input PDFs
(MRST~\cite{Martin:1998sq} with \asmz=0.1175, and
CTEQ5M~\cite{Lai:1999wy} with \asmz=0.118), gave very good
agreement. This is shown in the right panel of
fig.~\ref{fig:cdf95}, taken from an update of~\cite{Frixione:1997ma}. 
The crucial change between the two predictions was the change in the
value of the QCD coupling strength \as\ extracted from global PDF
fits. The fits used in the CDF 1995 publication,
MRSD0~\cite{Martin:1992as}, did not include HERA data and had \asmz=0.111,
signficantly lower than what we were getting from LEP, namely
$\as(m_Z)\sim 0.120$. This 10\% difference, when evolved to the low scales
of relevance to $b$ production, becomes much more significant,
especially because $b$ rates grow like $\as^2$.
This is shown more explicitly in fig.~\ref{fig:pdfevol}. The left
panel shows the ratio of the rates obtained by using post-HERA PDF
sets, normalized to the MRSD0 set used in the CDF 1995 comparison. The
right panel shows the integrated total cross section (for $\vert y
\vert < 1$) for several PDF sets, ordered versus the date of
release. One can notice a constant increase, with the most recent
sets being almost a factor of 2 higher than the older ones. Notice
that this increase is due by and large to the increased value of $\as$
returned by the PDF fits. Forcing  $\Lambda_{\sss QCD}$
to take the value consistent with LEP's $\as(m_Z)$, one would
have obtained for each PDF set the values corresponding to the crosses
in the plots. There the increase relative to the pre-HERA fit MRSD0
is significantly smaller.

\begin{figure}[t]
\includegraphics[height=.3\textheight, clip]{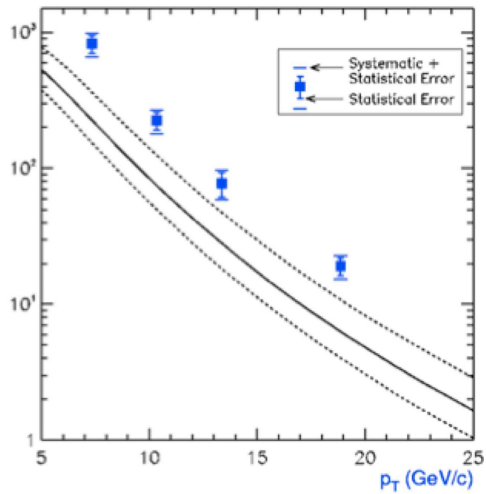}
\caption{Final CDF analysis of run~I exclusive-decay
  data~\cite{Acosta:2001rz}, compared to
  the CDF evaluation of the NLO QCD prediction with MRST
 PDFs and Peterson fragmentation.}
\label{fig:cdf02}
\end{figure}

While the improvements in the PDF fits were reducing the difference
between data and theory, as shown fig.~\ref{fig:cdf95}, a new CDF
measurement from the full sample of run~I exclusive $B$ decays
 in the range $6~\gev<\pt<20$~GeV appeared in
 2001~\cite{Acosta:2001rz}, and is shown here in Fig.~\ref{fig:cdf02}. 
The total rate turned out to be 50\% larger than in the
previous 1995
 publication~\cite{Abe:1995dv}: $\sigma(\pt(B^+)>6\gev, \vert
y\vert<1)=3.6\pm 0.6\mu$b, compared to the previous $2.4\pm
 0.5\mu$b, a change in excess of 2$\sigma$.  
 The ratio between data and the central value of the theory prediction
was quoted as $2.9\pm 0.5$: a serious disagreement was back!

\begin{figure}[t]
  \includegraphics[width=.23\textheight, angle=-90]{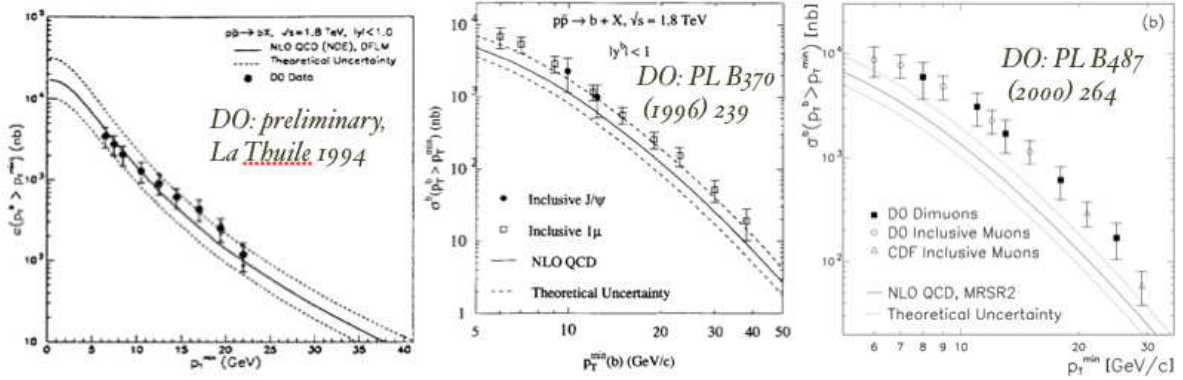}
\caption{Evolution of the D0 measurements. Left: preliminary results
  from 90nb$^{-1}$~\cite{Bazizi:1994sp}. Center:
  6.6pb$^{-1}$~\cite{Abachi:1996jq}. Right: final analysis of the same
  data set, with the addition of inclusive
  dimuons~\cite{Abbott:1999se}.}
\label{fig:D094-00}
\end{figure}

On the other side of the Tevatron ring, the D0 experiment started
presenting the first $b$ cross section measurements in 1994. 
The first preliminary results~\cite{Bazizi:1994sp} were in perfect
agreement with QCD, as shown in the left panel of
Fig.~\ref{fig:D094-00}. They were eventually published, after
significant changes, in~\cite{Abachi:1994kj}. The results from a
larger dataset of 6.6pb$^{-1}$ appeared in~\cite{Abachi:1996jq}, where
$\psi$ dimuons were added. They
are shown in the central panel of the figure, and they show a clear
increase over the preliminary analysis, but are still consistent with
the QCD expectations. The same data set underwent further analysis,
and eventually appeared few years later in~\cite{Abbott:1999se}. They
are shown in the right panel of the figure. Now
the data are significantly higher than QCD, and certainly higher than in
1996, especially in view of the fact that in the meantime the theory
predictions had increased by almost a factor of 2 as a result of the
use of new PDF sets (this is clearly visible by the shift of the
theory curves between the central and right panels). As in the case of the
CDF exclusive analysis, this evolution
underscores the difficulty in performing these measurements, and
indicates that it was not just the theory that was having difficulties
in coming to grips with the problem!

\begin{figure}[t]
  \includegraphics[width=.17\textheight, angle=-90]{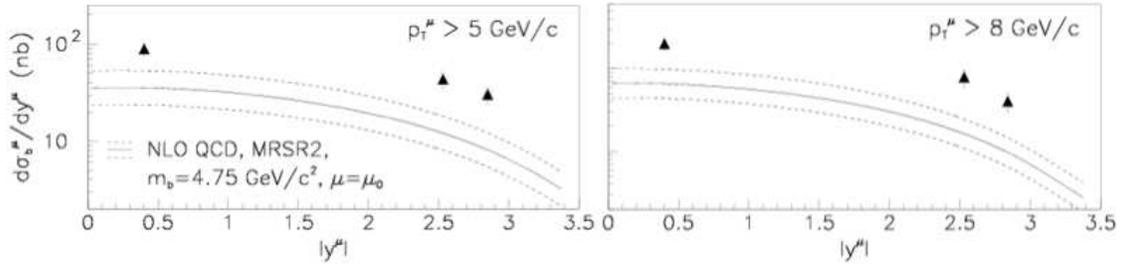}
\caption{Forward muon production at D0~\cite{Abbott:1999wu}.}
\label{fig:D0fwdmu}
\end{figure}

An additional element was added to the puzzle when D0
reported~\cite{Abbott:1999wu} the
measurement of $b$ production at large rapidity, using inclusive
forward muons ($2.4<\vert y_\mu \vert <3.2$). The results, shown in
fig.~\ref{fig:D0fwdmu}, indicated an excess over NLO QCD by a factor
larger than what observed in the central region. 
This anomaly could not be explained away by assuming some extra
systematics related to PDFs. From the point of view of perturbation
theory, furthermore, there was no reason to expect a significant
deterioration of the predictive power when going to large rapidity.
So when this result first appeared in its preliminary form I was
led~\cite{Mangano:1997ri} to review our assumptions about the
non-perturbative part of the calculation, in particular the impact of
the fragmentation function.  A crucial observation is that in hadronic
collisions the fragmentation function is probed in different ranges of
$z$ as we change rapidity. This is easily seen as follows. Let us
assume that the $b$ \pt\ spectrum takes the simplified form:
\be
\frac{d\sigma(b)}{d p_T} \sim \frac{1}{p_T^N} \; ,
\ee
where the slope $N$ will typically depend on rapidity, becoming larger at
higher $y_b$. 
The meson spectrum is then obtained via convolution with the
fragmentation function $f(z)$, leading to the simple result:
\be \frac{d\sigma(B)}{d P_T} \equiv \int \frac{dz}{z}\,
 \frac{d\sigma(b)}{d p_T}(p_T= P_T/z) = 
\int \frac{dz}{z}\, (\frac{z}{
  P_T})^N\, f(z) =f_N \, \frac{d\sigma(b)}{d P_T} \; ,
\ee
where $f_N$ is the $N$-th moment of $f(z)$. This means that a steeper
partonic spectrum selects higher moments. Since the index $N$ is
larger for forward production, a relative difference in $B$ production
rates in the forward/central regions could be explained by making the
fragmentation function harder, enhancing the larger moments
(which measure the large-$z$ behaviour of $f(z)$). 
A related observation is that $f(z)$ fits to $e^+e^-$ data are mostly
 driven by the value of the first moment $f_1$, which measures the
 average of the fragmentation variable $z$. It is therefore possible
 that different choices of $f(z)$, giving equivalent overall fits to  
$e^+e^-$, could make very different predictions for the higher moments
 of relevance to hadronic production (in this case $N$ is in the range
 4-6). 

\begin{figure}[t]
  \includegraphics[height=.3\textheight]{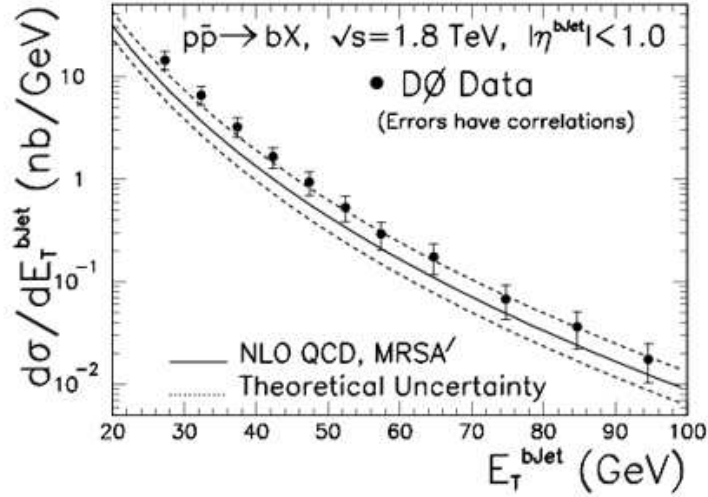}
\caption{$b$-jet production at D0~\cite{Abbott:2000iv}.}
\label{fig:D0bjet}
\end{figure}
One way to understand whether indeed the inaccurate description of the
fragmentation process could affect the theoretical predictions was
therefore to think of measurements not affected by this
systematics.
The most obvious observable of this kind is the \et\
spectrum of jets containing a $b$ quark~\cite{Frixione:1996nh}.  Since
the tagging of a $b$ inside the jet is only marginally affected by the
details of the $b\to B$ fragmentation, measuring the rate of $b$ jets
is a direct measurement of the $b$ production rate with negligible
fragmentation systematics. In addition, this measurement is also
insensitive to higher-order large-\pt\ logarithms which are present in
the \ptb\ spectrum, therefore improving in principle the perturbative
accuracy.  D0 carried out the measurement,
publishing~\cite{Abbott:2000iv} the results shown in
Fig.~\ref{fig:D0bjet}. The agreement with NLO
QCD~\cite{Frixione:1996nh} is better than in
the case of the \ptb\ spectrum, as was hoped. We took this as
strong evidence that a reappraisal of the fragmentation function
systematics may have led to a better description of the \ptb\ and
$y_\mu$ distributions. 

\begin{figure}[t]
\includegraphics[width=.3\textheight, angle=-90]{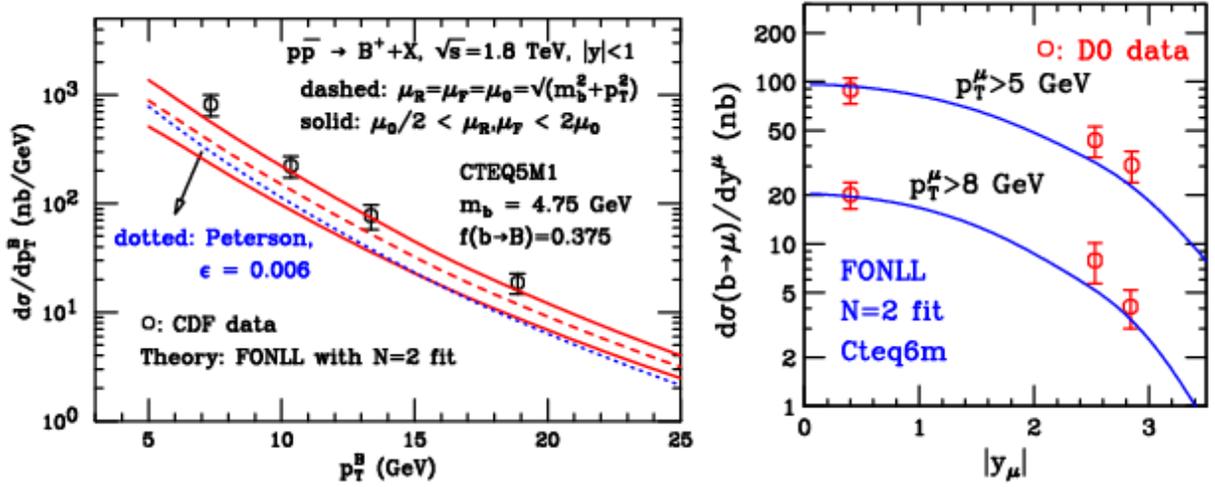}
\caption{Left panel: FONLL prediction by Cacciari and
  Nason~\cite{Cacciari:2002pa} for the run~I $B$ meson spectrum,
  compared to the CDF data~\cite{Acosta:2001rz}. Right panel: the
  prediction of this calculation for the forward muon rapidity
  spectrum at D0.}
\label{fig:cn02}
\end{figure}

The necessary ingredients to carry out this programme are
perturbative calculations of matching accuracy for $b$ spectra in
both $e^+e^-$ and $\ppbar$ collisions, in addition of course to
accurate $e^+e^-$ data to be used in the fits. These tools had just
become available towards the end of the 90's
The resummation of the logarithms of $\pt/m_b$, with next-to-leading
logarithmic accuracy (NLL), and the matching with the fixed-order (FO),
exact NLO calculation for massive quarks, had been performed in
\cite{Cacciari:1998it} (Fixed-Order with Next-to-Leading-Log
resummation: FONLL)
and a calculation with this level of accuracy for
 $e^+e^-$ collisions was presented
in~\cite{Nason:1999zj}. Here it had been used for the
extraction of the non-perturbative fragmentation 
function $f(z)$ from LEP and SLC
data~\cite{Heister:2001jg}, with the main result that the Peterson
 functional form is strongly disfavoured over other
 alternatives~\cite{Kartvelishvili:1977pi}.  
The equivalence of the perturbative
inputs allows one to consistently apply this fit to the
FONLL $b$-quark spectra in hadronic collisions, leading to FONLL
predictions for the $b$ hadron ($H_b$) spectrum.  A comparison of these
predictions with the final CDF data at 1.8~TeV for $B^\pm$-meson
 production in the range $6~\gev<\pt<20$~GeV has been presented
 in~\cite{Cacciari:2002pa}. The results are shown in
 Fig.~\ref{fig:cn02}: the left panel compares  the CDF data
 from~\cite{Acosta:2001rz} 
 with the theory curve evaluated using CTEQ5M PDF, FONLL, and
 fragmentation functions fitted to LEP and SLC data. 
The right panel shows a comparison~\cite{Nason:2003zf}
 with the D0 forward muon data. 
In both cases the agreement with data
 is much improved. In the case of the CDF central cross section,
 the ratio between data and theory improves from $2.9\pm 0.5$ to 
 $1.7\pm 0.7$. As discussed in detail in~\cite{Cacciari:2002pa}, the
 improvement is due to the sum of three independent 20\% effects
 ($1.2^3\sim 2.9/1.7$), all going in the same direction: 
the resummation of \pt\ logarithms, the change
 in functional form of the fragmentation function, and the use
 of the LEP/SLC $b$ fragmentation data.  
The heritage of run~I was therefore a set of measurements, more or
 less consistent with each other, normalized with a factor of about 1.5 to 2
higher than the central theoretical prediction, but still compatible
 with the upper end of the theoretical systematics band.

\section{The run~II CDF results}
The final phase of this history deals with the new run~II data from
CDF~\cite{cdfrun2}.  A great improvement took place in the ability to
trigger on very low \ptb\ events, allowing for a measurement down to
$\ptb\sim 0$, although still in the limited rapidity range $\vert y_b
\vert \lsim 0.6$. This is also accompanied by very large statistics,
allowing a fine binning in \pt.  The measurement down to very small
\ptb\ is important because the total rate has a much reduced
dependence on the fragmentation systematics, and because it is
particularly sensitive to possible small-$x$ phenomena.

On the theoretical side, in addition to the calculations described
above, a new tool has meanwhile become available, namely the 
MC@NLO code~\cite{Frixione:2003ei}, which
merges the full NLO matrix elements with the complete shower evolution
and hadronization performed by the {\small HERWIG} Monte Carlo. As discussed in
detail in~\cite{Frixione:2003ei}, this comparison probes a few features
where FONLL and MC@NLO differ by 
effects beyond NLO: the evaluation
of subleading logarithms in higher-order emissions, in particular in
the case of gluon emission from the $b$ quark, and the hadronization
of the heavy quark, which in MC@NLO is performed through {\small
HERWIG}'s cluster model, tuned on $Z^0\to H_b X$ decays.

\begin{figure}[t]
  \includegraphics[height=.3\textheight]{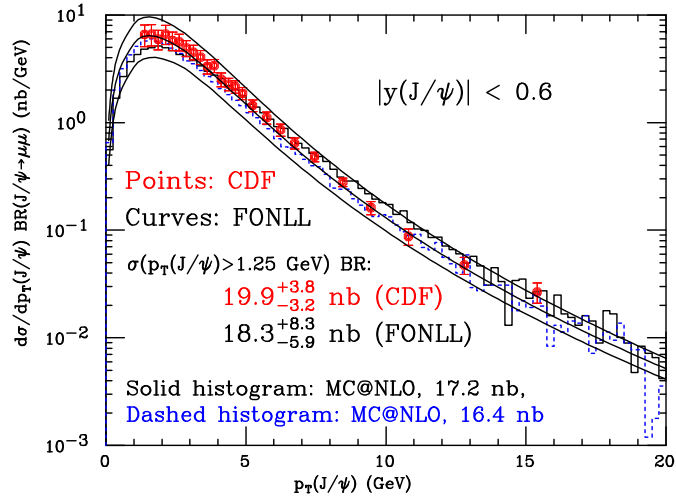}
\caption{CDF $J/\psi$ spectrum from $B$ decays. 
The theory band represents the FONLL systematic uncertainties,
as described in the text. Two MC@NLO predictions are
also shown (histograms).}
\label{fig:cdfpsi}
\end{figure}
 The comparison of the run~II data with the theoretical calculations
 is given in Fig.~\ref{fig:cdfpsi}, which shows
the data with our prediction for the
spectrum of $J/\psi$s form $H_b$ decays, 
obtained by convoluting the FONLL result with the
$J/\psi$ momentum distribution in inclusive $B\to J/\psi+X$
decays. The theoretical error band is obtained by varying 
renormalization and factorization scales ($\mu_{R,F}=\xi_{R,F}\mu_0$,
 with $\mu_0^2=\pt^2+m_b^2$), the $b$-quark mass, and parton densities.
The central values of our predictions are obtained with $\xi_{R,F}=1$,
$m_b=4.75$~GeV and CTEQ6M. The mass uncertainty corresponds to the range
$4.5~\gev<m_b<5$~GeV.
The scale uncertainty is obtained by varying $\mu_{R,F}$
over the range
 $0.5<\xi_{R,F}<2$, with the constraint $0.5 <\xi_R/\xi_F < 2$.
The PDF uncertainty is calculated by using all the three sets of PDFs with
 errors given by the CTEQ, MRST and Alekhin 
groups~\cite{Pumplin:2002vw,Martin:2002aw,Alekhin:2002fv}.

The data lie well within the uncertainty band, and are in very good
agreement with the central FONLL prediction. I also show the two
MC@NLO predictions corresponding to the two different choices of the
$b$ hadronization parameters (see~\cite{Cacciari:2003uh} for the
details).

\begin{figure}[t]
  \includegraphics[height=.3\textheight]{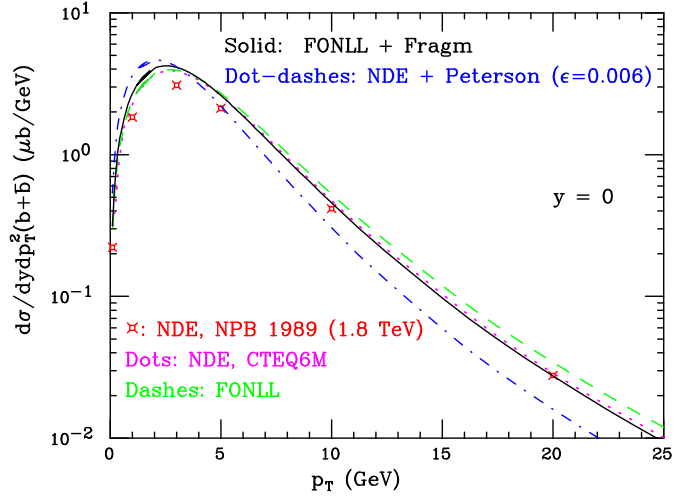}
\caption{Evolution of the NLO QCD predictions over time, for $\sqrt{S} =
1800$ GeV. See the text for the meaning of the various curves.}
\label{fig:history}
\end{figure}

I stress that both FONLL and MC@NLO are based on the NLO result
of~\cite{Nason:1989zy} (henceforth referred to as NDE), 
and only marginally enhance the cross section
predicted there, via some higher-order effects.  The most relevant
change in FONLL with respect to old predictions lies at the
non-perturbative level, i.e. in the treatment
of the $b\to H_b$ hadronization, which makes use~\cite{Cacciari:2002pa}
of the
moment-space analysis of the most up-to-date data on $b$ fragmentation
in $e^+e^-$ annihilation.  The evolution of the NLO theoretical
predictions over time is shown in Fig.~\ref{fig:history}. Here
we plot the original central prediction of NDE for $\sqrt{S}=$1.8~TeV 
(symbols), obtained using NLO QCD partonic cross sections convoluted with 
the PDF set available at the time, namely DFLM260~\cite{Diemoz:1987xu}. 
The same calculation, performed with the CTEQ6M
PDF set (dotted curve),
shows an increase of roughly 20\% in rate in the region
$\pt<10$~GeV. The effect of the inclusion of the resummation of NLL
logarithms is displayed by the dashed curve, and is seen to be modest
in the range of interest. Finally, we compare the
original NDE prediction after convolution with the Peterson
fragmentation function ($\epsilon=0.006$, dot-dashed curve), with the
FONLL curve convoluted with the fragmentation function extracted
in~\cite{Cacciari:2002pa} (solid curve). 
Notice that the effect of the fragmentation obtained in~\cite{Cacciari:2002pa} 
brings about a modest decrease of the cross section (the difference 
between the dashed and solid curves), 
while the traditional Peterson fragmentation with 
$\epsilon=0.006$ has a rather pronounced effect (the difference 
between the symbols and the dot-dashed curve).
Thus, the dominant change in the theoretical prediction 
for heavy flavour production from the original NDE calculation 
up to now appears to be the consequence of more precise 
experimental inputs to the bottom fragmentation
function~\cite{Heister:2001jg}, that have shown that non-perturbative 
fragmentation effects in bottom production are much smaller 
than previously thought.

\begin{figure}[t]
  \includegraphics[height=.3\textheight]{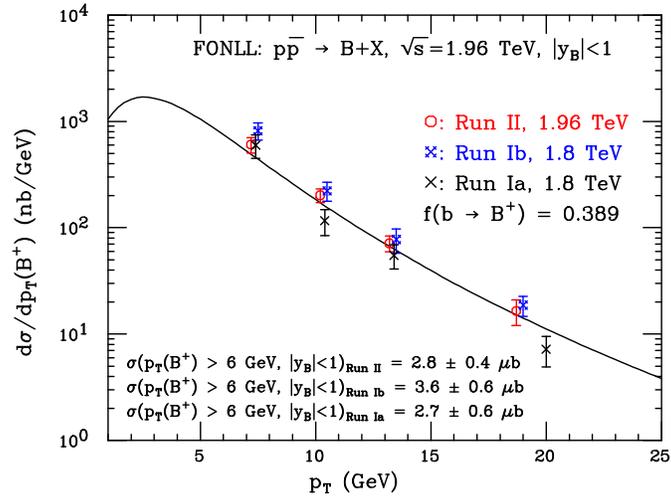}
\caption{Evolution of the CDF data for exclusive $B^\pm$ production:
  Run~IA\protect\cite{Abe:1995dv}, Run~IB~\protect\cite{Acosta:2001rz} and
  Run~II\protect\cite{cdfrun2}.}
\label{fig:bplus}
\end{figure}

The main improvement in the comparison between data and theory
w.r.t. the final run~I results discussed in~\cite{Cacciari:2002pa}
comes from the normalization of the run~II CDF data, which tend to be
lower than one would have extrapolated from the latest measurements at
1.8~TeV. To clarify this point, we collect in Fig.~\ref{fig:bplus} the
experimental results from the CDF measurements of the $B^\pm$ cross
section in Run~IA~\cite{Abe:1995dv}, in Run~IB~\cite{Acosta:2001rz}
and in Run~II. The rate for $\pt(B^\pm)>6$~GeV,
evolved from $2.4\pm0.5~\mub$ (Run~IA) to $3.6\pm0.6~\mub$ (Run~IB),
and decreased to $2.8\pm0.4~\mub$ in Run~II. The increase in the
c.m. energy should have instead led to an increase by 10-15\%. The
Run~II result is therefore lower than the extrapolation from Run~IB by
approximately 30\%. By itself, this result alone would reduce the
factor of 1.7 quoted in~\cite{Cacciari:2002pa} to 1.2 at $\sqrt{S} =
1.96$~TeV. In addition, the results presented
in~\cite{Cacciari:2003uh} lead to an increase in rate relative to the
calculation of~\cite{Cacciari:2002pa} by approximately 10-15\%, due to
the change of PDF from CTEQ5M to CTEQ6M.  We then conclude that the
improved agreement between the Run~II measurements and perturbative
QCD is mostly a consequence of improved experimental inputs (which
include up-to-date $\as$ and PDF determinations).

\section{Conclusions}
When I meet colleagues and discuss the latest $b$ results, and when
I hear presentations or read conference proceedings, there is
often a more or less explicit message that now things are OK
because theorists kept beating on their calculations until they got them
right. I hope that this note will dispel this prejudice. The history
of the experimental measurements indicates that many things have also
``strongly evolved'' on the data side, often with changes well in
excess of the standard $\pm 1\sigma$ variation. 
 The ``history'' plot in fig.~\ref{fig:history} shows on the other
 hand that not much
has changed on the theory side, aside from data-driven modifications
associated to the value of \asmz, to the low-$x$ behaviour of the
gluon as determined by the HERA data, and to the improved data on
$b\to B$ fragmentation. The theoretical improvements due to the
resummation of the large-\pt\ logarithms play a major role in allowing
a consistent use of the fragmentation functions extracted from
$e^+e^-$ data, but have a very limited impact in the region of \ptb\
probed by the run~II data. Their significance will only manifest
itself directly at
high \ptb\ ($\ptb>20-30$GeV), where the resummation leads to a much
reduced scale dependence, and to more accurate predictions, allowing
more compelling quantitative tests of the theory. It is auspicable
that the improved run~II detectors and the higher statistics will make it
possible to extend the range of the measurements to really large \ptb\
(in the range of 80-100 GeV). Tools are now available (MC@NLO) to
compare data subject to complex experimental constraints directly with
realistic NLO calculations, including a complete description of the
hadronic final state. This will avoid the risky business of attempting
to connect the observables to a \pt\ spectrum of the $b$ quark, a
practice which, although unavoidable in the past, has certainly
contributed to the inflation of theoretical and experimental
systematic uncertainties. 

To this date, the recent CDF measurement of total $b$-hadron
production rates in $p\bar{p}$ collisions at $\sqrt{S}=1.96$~TeV is in
good agreement with NLO QCD, the residual discrepancies being well
within the uncertainties due to the choice of scales and, to a lesser
extent, of mass and PDF. A similar conclusion is reached for the \pt\
spectrum.  The improvement in the quality of the agreement between
data and theory relative to previous studies is the result of several
small effects, ranging from a better knowledge of fragmentation and
structure functions and of \as, which constantly increased in the DIS
fits over the years, to the fact that these data appear to lead to
cross sections slightly lower than one would have extrapolated from
the measurements at 1.8~TeV.  The currently still large uncertainties in data
and theory leave room for new physics. However there is no evidence
now that their presence is required for the description of the data,
and furthermore the recent results of~\cite{Janot:2004cy} rule out the
existence of a scalar bottom quark in the range preferred by the
mechanism proposed in~\cite{Berger:2000mp}.  The data disfavour the
presence of small-$x$ effects of the size obtained with the approaches
of refs.~\cite{Levin:1991ry}.  They are instead compatible with the
estimates of~\cite{Collins:1991ty}.  

While these results have no direct impact on other anomalies
reported by CDF in the internal structure and correlations of
heavy-flavoured jets~\cite{Acosta:2001ct}, we do expect that the
improvements relative to pure parton-level calculations present in the
MC@NLO should provide a firmer benchmark for future studies of the
global final-state stucture of $b\bar{b}$ events.

\begin{theacknowledgments}
I am thankful to Harry Weerts and the other local organizers for
 the invitation and for the pleasant hospitality, and to Joey Huston
 for the support provided.
I am also deeply grateful to my friends, with whom I shared the challenge of
 understanding $b$ production properties over more than 10 years:
 M.Cacciari, S.Frixione, P.Nason and G.Ridolfi, plus all the pals and
 collaborators in CDF and D0.
\end{theacknowledgments}

\end{document}